\documentclass[10pt,english]{article}
\usepackage{amsthm}
\usepackage{amsmath}
\usepackage{amssymb}
\usepackage{setspace}

\begin{document}

\title{Classical probabilistic realization of ``Random Numbers Certified by Bell's Theorem''}

\author{Andrei Khrennikov\\
International Center for Mathematical Modeling\\
 in Physics and Cognitive Sciences \\
 Linnaeus University,  V\"axj\"o-Kalmar, Sweden}

\date{}

\maketitle

\begin{abstract} We question the commonly accepted statement that random numbers certified by Bell's theorem 
carry some special sort of randomness, so to say, quantum randomness or intrinsic randomness. We show that such numbers
can be easily generated by classical random generators. 
\end{abstract}

\section{Introduction}

The idea that quantum randomness (QR) differs crucially from classical randomness (CR) is due to von Neumann who 
pointed that QR is intrinsic and CR is reducible to variability in an ensemble \cite{VN}.
At the first stages of the development of quantum theory (QM) von Neumann's viewpoint to QR and its exceptional 
nature was merely of the purely foundational value. It was used to support various 
no-go statements, starting with von Neumann's no-go theorem \cite{VN} (in fact, this 
no-go statement was called by von Neumann ``ansatz'' and ``theorem'' as it is common in modern literature), see \cite{INT}, 
\cite{KHR-context} for discussions. 
However, with the development of quantum information theory and 
its technological implementation the inter-relation between classical and quantum randomness started to play 
the important role in real applications. 

In this note we shall discuss one very novel and important application of quantum randomness,
namely, in the form of random number generators certified by Bell's theory, see, e.g., \cite{77} for details and review. 
As was claimed in \cite{77}:

\medskip

 {\small ``It is thereby possible to design of a new
type of cryptographically secure random number generator which does not require any assumption
on the internal working of the devices. This strong form of randomness generation is impossible
classically and possible in quantum systems only if certified by a Bell inequality violation.''}

\medskip

It seems that this statement is in contradiction with the recent results \cite{KK}, see also \cite{KK0},  
on embedding of the data generated in experimental test
demonstrating violations of Bell's inequality into the classical probability model (based on the Kolmogorov measure-theoretic 
model \cite{K}, 1933). As well as in \cite{77}, we consider the case of CHSH inequality which were  tested and confirmed in numerous
experiments (although it might happen that the final loophole free test will be performed not for Bell's inequality not in  the CHSH  form, 
but in the Eberhard form \cite{BT1}).

\section{Classical probability space for data violating the CHSH inequality}  

For readers convenience, we present shortly the model from \cite{KK}. 

In the CHSH-test we operate with probabilities $p_{ij}(\epsilon,
\epsilon^\prime), \epsilon, \epsilon^\prime= \pm 1,$ -- to get the results
$a_{\theta_i}=\epsilon, b_{\theta_{j}^\prime}=\epsilon^\prime$ in
the experiment with the fixed pair of orientations $(\theta_i,
\theta_j^\prime).$  We borrow probabilities from the mathematical formalism of quantum mechanics:  
\begin{equation}
\label{QM} 
p_{ij}(\epsilon, \epsilon)=\frac{1}{2} \cos^2
\frac{\theta_i-\theta_j^\prime}{2}, p_{ij} (\epsilon,
-\epsilon)=\frac{1}{2} \sin^2 \frac{\theta_i-\theta_j^\prime}{2}.
\end{equation} 
 
Let us now consider the set of points $\Omega$ (the space of ``elementary events'' in Kolmogorov's terminology): 
$$
\Omega= \{(\epsilon_1, 0, \epsilon_1^\prime, 0), (\epsilon_1, 0, 0,
\epsilon_2^\prime), (0, \epsilon_2, \epsilon_1^\prime, 0), (0,
\epsilon_2, 0, \epsilon_2^\prime)\},
$$
where $\epsilon =\pm 1, \epsilon^\prime= \pm 1.$
 
We define the following probability measure\footnote{To match completely with Kolmogorov's terminology, we have to select a $\sigma$-algebra ${\cal F}$ of subsets of $\Omega$ representing events and define 
the probability measure on ${\cal F}.$ However, in the case of a finite $\Omega=\{\omega_1,..., \omega_k\}$ the system of events ${\cal F}$ is always chosen as consisting 
of all subsets of $\Omega.$ To define a probability measure on such ${\cal F},$ it is sufficient to define it for one-point sets, 
$(\omega_m) \to {\bf P}(\omega_m), \sum_m {\bf P}(\omega_m)=1,$ and to extend it by additivity:
 for any subset $O$ of $\Omega, {\bf P}(O)= \sum_{\omega_m\in O} 
{\bf P}(\omega_m).$  
}
on $\Omega$
$${\bf P}(\epsilon_1, 0, \epsilon_1^\prime, 0) = \frac{1}{4} p_{11}(\epsilon_1, \epsilon_1^\prime),
{\bf P}(\epsilon_1, 0, 0, \epsilon_2^\prime) = \frac{1}{4}
p_{12}(\epsilon_1, \epsilon_2^\prime)
$$
$$
{\bf P}(0, \epsilon_2, \epsilon_1^\prime, 0) = \frac{1}{4}
p_{21}(\epsilon_2, \epsilon_1^\prime), {\bf P}(0, \epsilon_2, 0,
\epsilon_2^\prime) = \frac{1}{4} p_{22}(\epsilon_2,
\epsilon_2^\prime).
$$

We now define random variables $A^{(i)}(\omega), B^{(j)}
(\omega):$
\begin{equation}
\label{QMA} 
A^{(1)}(\epsilon_1, 0, \epsilon_1^\prime, 0)= A^{(1)}(\epsilon_1,
0, 0, \epsilon_2^\prime)= \epsilon_1,
A^{(2)}(0, \epsilon_2, \epsilon_1^\prime, 0)= A^{(2)}(0, \epsilon_2, 0, \epsilon_2^\prime)= \epsilon_2;
\end{equation}
\begin{equation}
\label{QMB} 
B^{(1)}(\epsilon_1, 0, \epsilon_1^\prime, 0)= B^{(1)}(0,
\epsilon_2, \epsilon_1^\prime, 0)= \epsilon_1^\prime,
B^{(2)}(\epsilon_1, 0, 0, \epsilon_2^\prime)= B^{(2)}(0,
\epsilon_2, 0, \epsilon_2^\prime)=\epsilon_2^\prime.
\end{equation}
and we put these variables equal to zero in other points. 

\medskip

We find two dimensional probabilities $${\bf P}(\omega \in \Omega:
A^{(1)} (\omega)= \epsilon_1, B^{(1)}(\omega)= \epsilon^\prime_1)
={\bf P}(\epsilon_1, 0, \epsilon_1^\prime, 0)= \frac{1}{4} p_{11}
(\epsilon_1, \epsilon_1^\prime), \ldots,
$$
$$
{\bf P} (\omega \in \Omega: A^{(2)} (\omega)= \epsilon_2,
B^{(2)}(\omega)= \epsilon_2^\prime) = {\bf P}(0, \epsilon_2, 0,
\epsilon_2^\prime) =\frac{1}{4} p_{22}
(\epsilon_2, \epsilon_2^\prime).
$$

We also consider the random variables which are responsible for
selections of pairs settings $(\theta_i, \theta_j^\prime).$ For ``the settings at LHS'': 
$$
\eta_L(\epsilon_1, 0, 0,\epsilon_2^\prime)= \eta_L(\epsilon_1, 0, \epsilon_1^\prime, 0)= 1,
\eta_L(0, \epsilon_2, 0, \epsilon_2^\prime)= \eta_L(0, \epsilon_2, \epsilon_1^\prime, 0)= 2 .
$$
For ``the settings at RHS'': 
$$
 \eta_R(\epsilon_1, 0, \epsilon_1^\prime, 0)= \eta_R(0, \epsilon_2,\epsilon_1^\prime, 0)= 1, 
\eta_R(0, \epsilon_2, 0, \epsilon_2^\prime)=  \eta_R(\epsilon_1, 0, 0,\epsilon_2^\prime)= 2.
$$
 
The points of the space of ``elementary events'' $\Omega$ have the following interpretation. 
Take,for example, $\omega=(\epsilon_1, 0, \epsilon_1^\prime, 0).$ Here $\eta_L=1,$ i.e, the LHS
setting is selected as $\theta_1,$ and    $\eta_R=1,$ i.e, the RHS
setting is selected as $\theta_1^\prime.$ Then the random variable $A^{(1)}=\epsilon_1, A^{(1)}=0$
and $B^{(1)}= \epsilon_1^\prime, B^{(2)}=0.$ Consider now, e.g., $\omega= (\epsilon_1, 0, 0,
\epsilon_2^\prime).$ Here $\eta_L=1,$ i.e, the LHS
setting is selected as $\theta_1,$ and    $\eta_R=2,$ i.e, the RHS
setting is selected as $\theta_2^\prime.$ Then the random variable $A^{(1)}=\epsilon_1, A^{(1)}=0$
and $B^{(1)}= 0, B^{(2)}=\epsilon_2^\prime.$

\section{Classical random generation of ``Random Numbers Certified by Bell's Theorem''}

By using the previous classical probabilistic representation of CHSH-probabilities,  
we can  now construct the classical random generator (at least theoretically) producing 6-dimensional
vectors $\xi_m=(A^{(1)}_m, A^{(2)}_m, B^{(1)}_m, B^{(2)}_m, \eta_{Lm}, \eta_{Rm}), m=1,2,....$ The last two coordinates 
take values 1,2 and the first four coordinates take the values $\{-1, 0, +1\}.$ Now if we filter from these
6-dimensional vectors the zero coordinates, we shall get 4-dimensional vectors $(a_m, b_m, \eta_{Lm}, \eta_{Rm}$ 
which are indistinguishable from the viewpoint of Bell's test from those obtained, e.g.,  in the quantum optics 
experiments. The corresponding conditional probabilities, conditioning with respect to selection the fixed pairs of 
settings $(\theta_i, \theta_j^\prime),$ violate the CHSH inequality. 

Finally we remark that our random generator is ``local'', see \cite{KK0} for details.   
 
 \section{Conclusion}
 
In the light of recent author's results on embedding of quantum probabilistic data into the classical Kolmogrov model of probability
von Neumann's statement about irreducible and intrinsic quantum randomness 
as opposed to classical reducible randomness which based only on the lack 
of knowledge has to be seriously  reanalyzed (as well as its quantum information technological implementations). 
In fact, already careful reading of von Neumann's book \cite{VN} makes the impression that his conclusion about novel features
of quantum probability and randomness comparing with their classical counter-parts was not completely logically justified. Roughly speaking 
he said. See, each electron is individually random! However, at the same time he pointed out that this individual randomness can be gained 
only from an ensemble of electrons. The latter ensemble approach to probability is purely classical. Therefore it would be natural to expect 
that one might be able to embed quantum probability into the classical probability model. Precisely this was done in \cite{KK}, \cite{KK0}.

In our approach the CHSH-probabilities appear as {\it classical conditional probabilities.} In general I claim that all quantum probabilities 
can be modeled as classical conditional probabilities. In such a situation it is difficult to believe in some mystical and non-classical 
features of quantum random generators.

Finally, I remark that in this paper I do not question the quality of ``Random Numbers Certified by Bell's Theorem''. If they pass the standard 
tests for randomness, e.g., the NIST-test, they are good, if not, then they are bad, irrelative to non/violation of the CHSH-inequality and irrelative 
to the problem of closing of experimental loopholes \cite{BT1}. My main concern is about claims that quantum random generators are in some way better 
than classical ones. 

\section*{Acknowledgment}

This paper was written under support of the grant Modeling of Complex Hierarchic Systems, the Faculty of technology, Linnaeus University.


\begin{thebibliography}{400}

\bibitem{VN}  J. Von Neuman,  Mathematical Foundations of Quantum Mechanics. Princeton University Press, Princeton (1955).

\bibitem{INT}   A. Khrennikov, {\it Interpretations of Probability} (Berlin: De Gruyter) 2d ed (2010).

\bibitem{KHR-context} A. Khrennikov   Contextual Approach to Quantum Formalism.
Springer, Berlin-Heidelberg-New York (2009).

\bibitem{77} S. Pironio, A. Acin, S. Massar, A. Boyer de la Giroday, D. N. Matsukevich, P. Maunz, S. Olmschenk, D. Hayes, L. Luo, T. A. Manning, C. Monroe
Random Numbers Certified by Bell's Theorem. 	Nature 464, 1021 (2010).

  \bibitem{KK} Avis D, Fischer P, Hilbert A, and Khrennikov A 2009 Single, Complete, 
Probability Spaces Consistent With EPR-Bohm-Bell Experimental Data, 
{\it Foundations of Probability and Physics-5}
 vol 750 (Melville, NY: AIP)  pp  294-301.
 
\bibitem{KK0} A.   Khrennikov 2014 Classical probability model for Bell inequality.
{\it EmQM13: Emergent Quantum Mechanics3–6 October 2013, Vienna, Austria.}
J. Phys.: Conf. Ser., 504.

\bibitem{K} Kolmogoroff A N 1933 {\it  Grundbegriffe der Wahrscheinlichkeitsrechnung} ( Berlin: Springer Verlag);
English translation: Kolmogorov A N 1956 {\it Foundations of Theory of
Probability} (New York: Chelsea Publishing Company)


\bibitem{BT1}  Giustina M, Mech Al,  Ramelow S, Wittmann B, Kofler J,  Beyer J, Lita A, Calkins B, 
Gerrits Th, Woo Nam S,  Ursin R, and  Zeilinger A  2013 {\it Nature} {\bf 497} 227  


\end{thebibliography}
\end{document}